\documentclass[preprint,12pt]{elsarticle}

\usepackage{graphics}
\usepackage{graphicx}
\usepackage{epsfig}
\usepackage{amssymb}

\journal{Applied Radiation and Isotopes}

\begin{document}

\begin{frontmatter}

\title{Half-life measurement of $^{66}$Ga with $\gamma$-spectroscopy}

\author[ATOMKI]{Gy. Gy\"urky\corref{cor}}
\ead{gyurky@atomki.hu}
\author[ATOMKI]{J. Farkas}
\author[ATOMKI]{Z. Hal\'asz}
\author[ATOMKI]{T. Sz\"ucs}
\address[ATOMKI]{Institute of Nuclear Research (ATOMKI), H-4001 Debrecen, POB.51., Hungary}
\cortext[cor]{corresponding author}
\begin{abstract}

The half-life of $^{66}$Ga, an isotope very important for high-energy efficiency calibration of $\gamma$-detectors, has been measured using $\gamma$-spectroscopy. In order to reduce systematic uncertainties, different source production methods and $\gamma$-counting conditions have been applied. A half-life value of \linebreak t$_{1/2}$\,=\,(9.312\,$\pm$\,0.032)\,h has been obtained in agreement with a recent measurement but in contradiction with some of the earlier results.

\end{abstract}

\begin{keyword}

$^{66}$Ga  \sep half-life \sep $\gamma$-spectroscopy

\end{keyword}

\end{frontmatter}

\section{Introduction}
\label{sec:introduction}

$^{66}$Ga is a very important isotope for the efficiency calibration of $\gamma$-detectors since it emits $\gamma$-rays with energies up to almost 5\,MeV. It has a relatively short half-life of less than 10 hours, therefore, it is not available as a commercial calibration standard, thus it needs to be produced practically in-situ for a detector efficiency measurement. For an absolute efficiency calibration the source activity must be known with high precision which necessitates the precise knowledge of the decay half-life. In 2004 a critical review has been published about the half-lives of radionuclides considered to be important for detector efficiency calibrations \cite{woo04}. It has been found that the precision of the $^{66}$Ga half-life is by far not enough for the requirements posed by the International Atomic Energy Agency \cite{IAEA91}. Therefore new half-life measurements of $^{66}$Ga is recommended by Ref.\,\cite{woo04}.

Since 2004 two new high precision half-life measurement of $^{66}$Ga became available whose results disagree by about six standard deviations. $\gamma$-spectroscopy measurement of K. Abbas \textit{et al.} gives t$_{1/2}$\,=\,(9.49\,$\pm$\,0.03)\,h \cite{abb06}, while based on positron detection G. W. Severin \textit{et al.} reports on a significantly lower value of t$_{1/2}$\,=\,(9.304\,$\pm$\,0.008)\,h \cite{sev10}. This strong deviation indicates that the knowledge of the $^{66}$Ga half-life is still very far from the required precision, therefore, new experiments are clearly needed. (For a complete list of the measured half-life values of $^{66}$Ga see Table\,1. of Ref.\,\cite{sev10}.)

In the present work the half-life of $^{66}$Ga has been measured based on counting the $\gamma$-radiation following the $\beta^+$ decay. In section \ref{sec:experimental} details of the experiments are discussed while the data analysis and the results are presented in section \ref{sec:results}.

\section{Experimental procedure}
\label{sec:experimental}

\subsection{Source preparation}

The isotope $^{66}$Ga can be produced by various nuclear reactions. In order to avoid any hidden systematic uncertainty on the deduced half-life coming from the source preparation procedure, two different reactions have been used to prepare the sources. The first set of sources has been produced by the $^{66}$Zn(p,n)$^{66}$Ga reaction. Natural isotopic composition zinc targets have been prepared by evaporating metallic zinc onto thick Al backing. In order to avoid any losses of the target layer during the half-life measurement, a thin protective Al layer has been evaporated onto each targets. The targets have been irradiated with 11.3\,MeV protons from the cyclotron accelerator of ATOMKI. Target current was typically 1\,$\mu$A and the irradiations lasted for about 30 minutes.

The second set of sources has been produced by the $^{63}$Cu($\alpha$,n)$^{66}$Ga reaction. Thick copper discs have been irradiated with 12.5\,MeV alphas from the cyclotron. Only one minute irradiation with 1\,$\mu$A intensity was enough to achieve high enough source activity. 

Besides $^{66}$Ga, other gallium isotopes have also been produced in the targets. In order to avoid the disturbing activity of the short lived reaction products (especially $^{64}$Ga, $^{65}$Ga and $^{68}$Ga), a cooling time of at least 6 hours has been inserted between the irradiations and the start of $\gamma$-countings. $^{67}$Ga has a longer half-life of 78.3\,h, therefore, its presence in the sources could not be avoided by cooling and its $\gamma$-lines were always present in the spectra. However, these $\gamma$-lines are mainly at low energies, not contributing to the Compton background at energies of interest for $^{66}$Ga. Besides the Ga isotopes, no other radioisotopes have been produced on the Cu targets in any significant amount. The backing of the Zn targets, however, contained some Fe impurity leading to the production of $^{56}$Co through the $^{56}$Fe(p,n)$^{56}$Co reaction. This isotope emits a high intensity $\gamma$-line at 1037.8\,keV which is very close to the energy of the most intense line of $^{66}$Ga (1039.2\,keV). Therefore, the 1039.2\,keV line could not be used for the half-life analysis in the case of the Zn sources.

With both production techniques three sources have been prepared. In the following they will be referred to as Zn-1, Zn-2, Zn-3, Cu-1, Cu-2 and Cu-3.

\subsection{Gamma-detection}

\begin{figure}
\centering
\resizebox{0.8\textwidth}{!}{\rotatebox{270}{\includegraphics{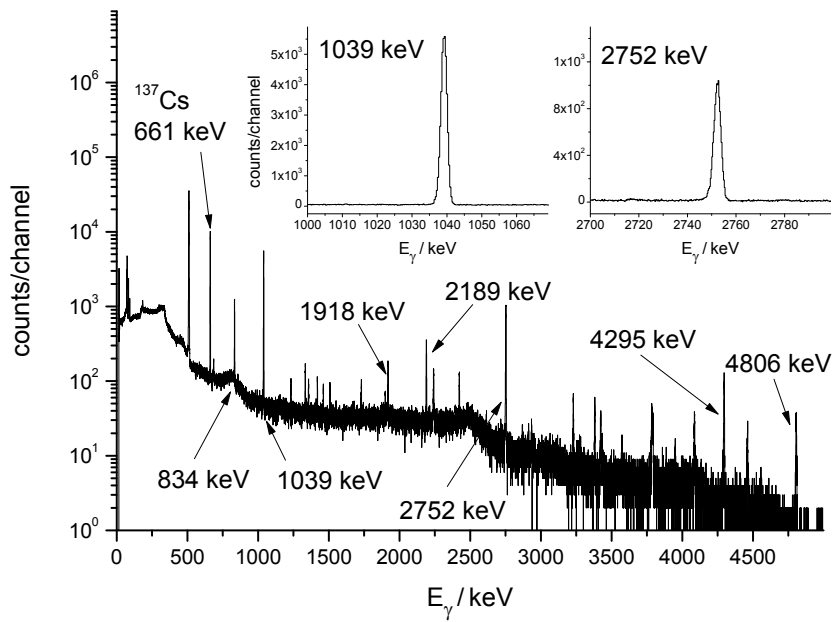}}}
\caption{\label{fig:spectrum} $\gamma$-spectrum taken on the Cu-1 source 10 hours after the start of counting for half an hour. The $^{66}$Ga peaks used for the analysis as well as the peak of the $^{137}$Cs reference source are indicated by arrows. The insets show two of the analyzed peaks.}
\end{figure}

Three separate counting setups have been used to detect the $\gamma$-radiation following the $\beta^+$ decay of $^{66}$Ga. Sources Zn-1 and Cu-1 have been measured with a 100\,\% relative efficiency Canberra model GR10024 HPGe detector with 7915-30-ULB cryostat configuration and equipped with a 10\,cm thick complete 4\,$\pi$ low background lead shielding. The measurements of the Zn-2 and Cu-2 sources have been carried out with a similar 100\,\% relative efficiency Canberra model GR10024 HPGe detector having a different cryostat configuration (7600SL) and 5\,cm thick lead shielding. The third set of sources have been measured with a 40\,\% relative efficiency Canberra model GR4025 HPGe detector without any shielding. The measurements with the three detectors have been running parallelly, however, completely independent nuclear electronics and data acquisition systems were used for the three detectors and they were placed in different counting rooms, therefore no cross-talk between the detectors could occur.

After signal processing the $\gamma$-spectra were recorded with ORTEC Model ASPEC-927 multichannel analyzers using the ORTEC MAESTRO software. In order to minimize the systematic uncertainty from dead time determination (see below), the measurements have been started when the dead time was below 2\,\%. The spectra were recorded in every half an hour until no count from $^{66}$Ga could be observed anymore. The total length of the counting varied between 45 and 87 hours. Figure \ref{fig:spectrum} shows a typical $\gamma$-spectrum taken on the Cu-1 source 10 hours after the start of counting. The insets show some of the peaks used for the analysis. The peak of the $^{137}$Cs reference source (see Sec. \ref{sec:systunc}) is also indicated.

$^{66}$Ga has a large number of $\gamma$-lines. In order to increase the reliability of the half-life determination, several stronger $\gamma$-lines have been used. These include the following (in parentheses the relative intensities are given): 833.5\,keV (5.9\,\%), 1039.2\,keV (37.0\,\%), 1918.3\,keV (2.0\,\%), 2189.6\,keV (5.3\,\%), 2751.8\,keV (22.7\,\%), 4295.2\,keV (3.8\,\%) and 4806.0\,keV (1.9\,\%) \cite{NDS10}.

\section{Data analysis and results}
\label{sec:results}

\begin{figure}
\centering
\resizebox{0.8\textwidth}{!}{\rotatebox{270}{\includegraphics{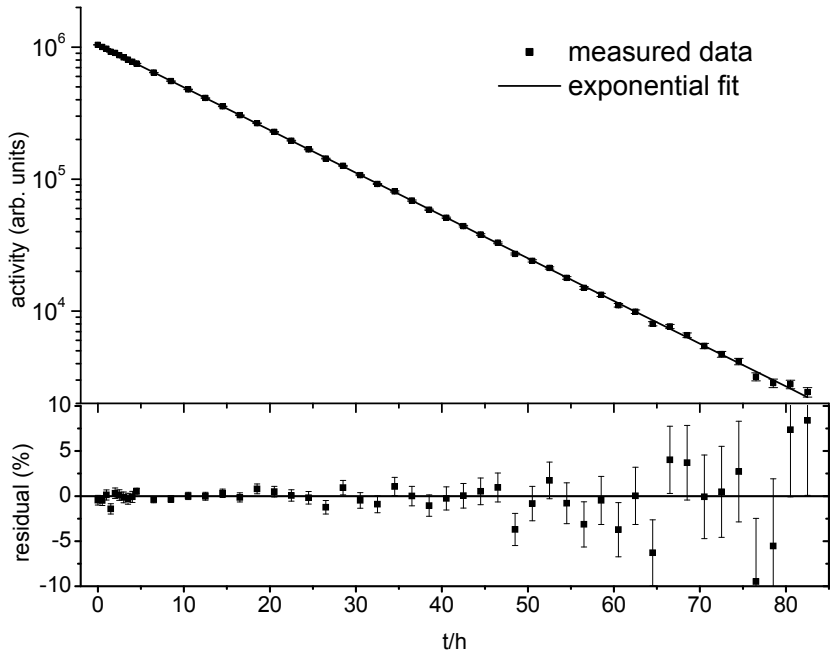}}}
\caption{\label{fig:decay} Decay curve based on the 1039\,keV peak of the Cu-2 sample.}
\end{figure}

The net peak areas have been determined with peak integration carried out with a small computer code developed for this measurement. The program allows for a manual selection of two background regions on the left and right side of the peaks. Linear background is fitted to these regions and based on this fit the background is subtracted from the manually selected peak region using proper error propagation. The half-life values have been determined for all samples and peaks separately by fitting an exponential function to the determined net peak areas as a function of time. The least square curve fit was done analytically following the procedure described e.g. in Ref.\,\cite{leo94}, the half-life value and its statistical uncertainty being determined from the fit parameter. A typical decay curve can be seen in Fig.\,\ref{fig:decay} where the activity of the Cu-2 sample based on the 1039\,keV peak is plotted as a function of time along with the least square fit.

Table \ref{tab:results} shows the obtained half-life results for all the studied samples and transitions. In the last row the $\chi^2_{\mathrm{red}}$ value and the degrees of freedom refer to the averaging while otherwise they refer to the exponential fit. The quoted uncertainties are statistical only. As one can see, there is a very good agreement within the dataset which is also indicated by the low $\chi^2_{\mathrm{red}}$ value of the average. Note, that in the average the results obtained with the GENIE data acquisition system (see below) are not taken into account. 

\begin{table}
\begin{footnotesize}
\centering
\caption{\label{tab:results} Half-life results. The given uncertainties are statistical only. See text for details.}
\begin{tabular}{lrr@{\,}c@{\,}lcc}
\hline
Source & E$_\gamma$/keV & \multicolumn{3}{c}{t$_{1/2}$/h} & d.o.f. & $\chi^2_{\mathrm{red}}$ \\
\hline
&	834	&	9.419	&	$\pm$	&	0.083	&	113	&	1.58	\\
&	1918	&	9.285	&	$\pm$	&	0.177	&	105	&	1.15	\\
Zn-1 &	2190	&	9.398	&	$\pm$	&	0.075	&	113	&	1.14	\\
&	2752	&	9.320	&	$\pm$	&	0.025	&	113	&	0.62	\\
&	4295	&	9.351	&	$\pm$	&	0.044	&	113	&	0.88	\\
\hline
&	834	&	9.408	&	$\pm$	&	0.046	&	114	&	0.89	\\
&	1918	&	9.282	&	$\pm$	&	0.111	&	114	&	0.66	\\
Zn-2 &	2190	&	9.265	&	$\pm$	&	0.060	&	114	&	1.09	\\
&	2752	&	9.334	&	$\pm$	&	0.021	&	114	&	1.00	\\
&	4295	&	9.374	&	$\pm$	&	0.055	&	114	&	1.06	\\
&	4806	&	9.464	&	$\pm$	&	0.093	&	114	&	0.77	\\
\hline
&	834	&	9.250	&	$\pm$	&	0.046	&	102	&	1.20	\\
&	1918	&	9.310	&	$\pm$	&	0.087	&	97	&	0.78	\\
Zn-3 &	2190	&	9.229	&	$\pm$	&	0.049	&	94	&	1.05	\\
&	2752	&	9.326	&	$\pm$	&	0.017	&	102	&	1.03	\\
&	4295	&	9.293	&	$\pm$	&	0.039	&	102	&	1.05	\\
&	4806	&	9.253	&	$\pm$	&	0.066	&	102	&	1.02	\\
\hline
&	834	&	9.316	&	$\pm$	&	0.073	&	43	&	1.19	\\
&	1039	&	9.321	&	$\pm$	&	0.024	&	43	&	0.91	\\
Cu-1 &	1918	&	9.319	&	$\pm$	&	0.210	&	43	&	0.50	\\
&	2190	&	9.393	&	$\pm$	&	0.105	&	43	&	0.68	\\
&	2752	&	9.310	&	$\pm$	&	0.043	&	43	&	0.97	\\
&	4295	&	9.483	&	$\pm$	&	0.301	&	43	&	0.73	\\
\hline
&	834	&	9.326	&	$\pm$	&	0.049	&	90	&	1.37	\\
&	1039	&	9.303	&	$\pm$	&	0.014	&	90	&	1.39	\\
&	1918	&	9.364	&	$\pm$	&	0.116	&	90	&	0.89	\\
Cu-2 &	2190	&	9.288	&	$\pm$	&	0.068	&	90	&	1.18	\\
&	2752	&	9.288	&	$\pm$	&	0.021	&	90	&	0.68	\\
&	4295	&	9.325	&	$\pm$	&	0.062	&	90	&	1.36	\\
&	4806	&	9.356	&	$\pm$	&	0.091	&	90	&	1.04	\\
\hline
&	834	&	9.372	&	$\pm$	&	0.034	&	100	&	1.16	\\
&	1039	&	9.298	&	$\pm$	&	0.011	&	100	&	0.89	\\
&	1918	&	9.407	&	$\pm$	&	0.077	&	100	&	0.90	\\
Cu-3 &	2190	&	9.222	&	$\pm$	&	0.052	&	97	&	1.30	\\
&	2752	&	9.297	&	$\pm$	&	0.019	&	100	&	1.05	\\
&	4295	&	9.370	&	$\pm$	&	0.053	&	100	&	1.11	\\
&	4806	&	9.306	&	$\pm$	&	0.084	&	100	&	0.97	\\
\hline
&	834	&	9.273	&	$\pm$	&	0.059	&	79	&	0.85	\\
&	1918	&	9.292	&	$\pm$	&	0.123	&	79	&	0.81	\\
Zn-3 &	2190	&	9.276	&	$\pm$	&	0.067	&	79	&	1.02	\\
GENIE &	2752	&	9.378	&	$\pm$	&	0.027	&	79	&	1.19	\\
&	4295	&	9.239	&	$\pm$	&	0.062	&	79	&	0.89	\\
&	4806	&	9.233	&	$\pm$	&	0.106	&	79	&	1.02	\\
\hline
\multicolumn{2}{l}{error weighted mean} & 9.312 &	$\pm$ & 0.005 & 36 & 0.98 \\
\hline
\end{tabular}
\end{footnotesize}
\end{table}

\begin{figure}
\centering
\resizebox{0.8\textwidth}{!}{\rotatebox{270}{\includegraphics{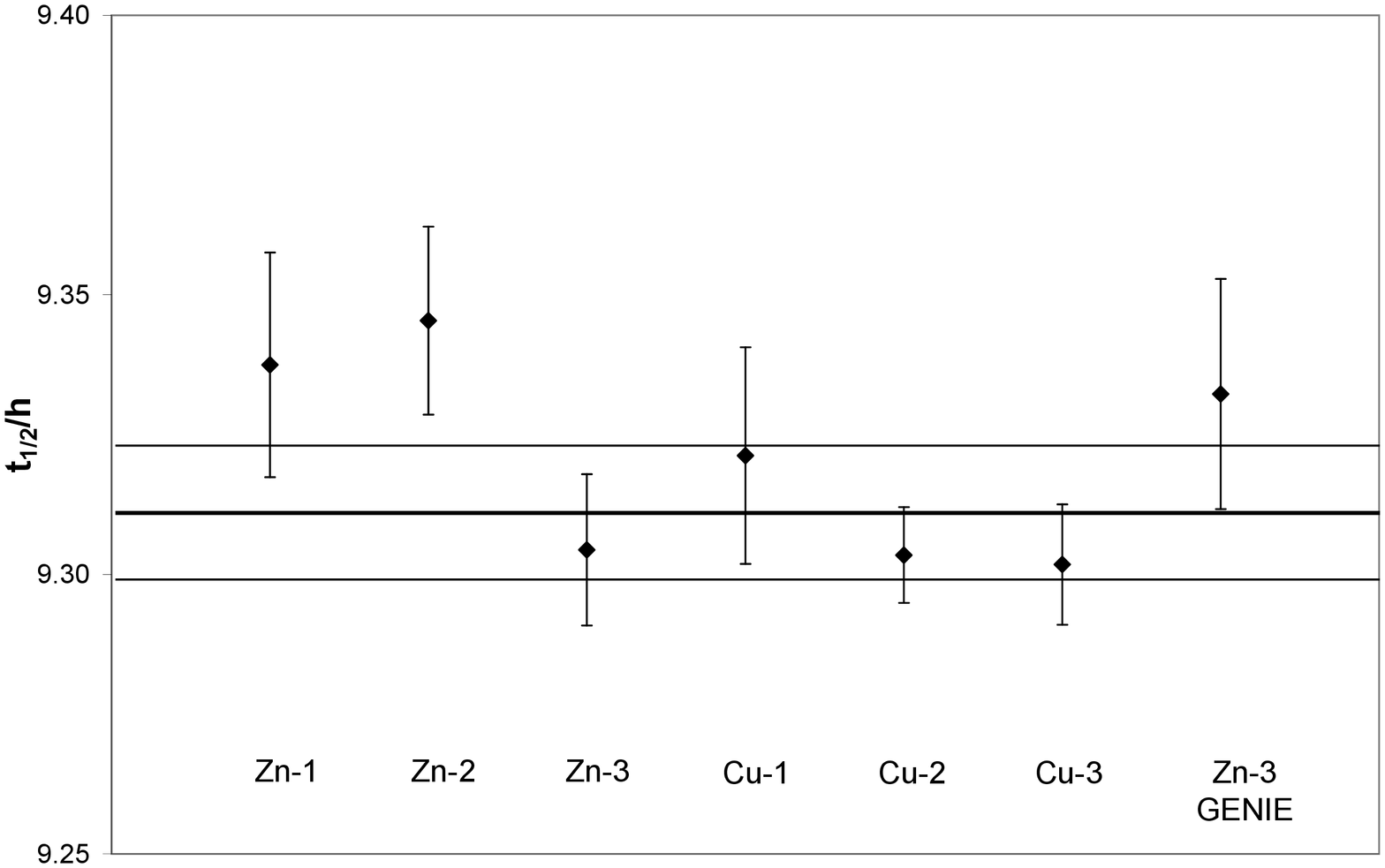}}}
\caption{\label{fig:results} Obtained half-life values for the six measured sources. The plotted values are the weighted averages of the half-lives determined from the analysis of the different $\gamma$-transitions. The horizontal lines indicate the weighted average of all the measurements and its two sigma statistical uncertainty.}
\end{figure}

\subsection{Systematic uncertainties}
\label{sec:systunc}

The careful assessment of uncertainties is of crucial importance for a reliable half-life determination \cite{pom08}. Owing to the fact that six sources have been measured, two different procedures have been used for the source production and three independent counting setups have been used for data taking, possible systematic uncertainties can be assessed by comparing the results obtained on the different sources. This comparison can be seen in Fig.\,\ref{fig:results} where the values assigned to the different sources have been obtained by taking the weighted averages of the measurements of several $\gamma$-transitions on a given sample as listed in Table \ref{tab:results}. The horizontal lines show the average value with its two sigma uncertainty. The $\chi^2_{\mathrm{red}}$ value of the average of the different sources is 1.60, somewhat higher than 1, but still acceptable for 5 degrees of freedom. Nevertheless, the difference between the half-life values obtained on different samples is used to estimate systematic uncertainties. 

Zn sources (and similarly Cu sources) have been prepared with the same technique, introducing possible correlation between their results. To account for this effect, results obtained on Zn and Cu samples are compared. The weighted average of the three Zn samples gives a half-life value of t$_{1/2}$(Zn)\,=\,(9.324\,$\pm$\,0.009)\,h while the same value for the Cu samples is \linebreak t$_{1/2}$(Zn)\,=\,(9.305\,$\pm$\,0.006)\,h. The standard deviation of these two values is 0.014\,h and this is assigned to the systematic uncertainty from source preparation. Similarly, the averages of the pairs of measurements carried out with the three detectors are calculated and the values obtained are: \linebreak t$_{1/2}$(Det--1)\,=\,(9.329\,$\pm$\,0.014)\,h, t$_{1/2}$(Det--2)\,=\,(9.312\,$\pm$\,0.008)\,h, \linebreak t$_{1/2}$(Det--3)\,=\,(9.303\,$\pm$\,0.008)\,h. The standard deviation of these values is 0.013\,h which is used as the systematic uncertainty of the different detectors used.

There are two common features of the six measurements. The first is that the same procedure is used to determine the net peak areas. To assess the uncertainty introduced by the peak integration method, different background and peak regions have been selected and the obtained peak areas have been used to derive half-life values. The obtained half-lives were always in agreement well within the given statistical uncertainties indicating the robustness of the net peak area determination. Typical differences of about 0.015\,h have been observed and this value is assigned to the systematic uncertainty of net peak area determination.

The other common feature of the six measurement is that in all cases the ORTEC MAESTRO data acquisition system has been used and the dead time corrections have been applied based on the real and live time values provided by the system. An error of the dead time values influence all measurements in the same way leading to a possible common source of systematic uncertainty, therefore a careful check of the reliability of the dead time is of great importance. 

As mentioned before, the $\gamma$-countings have been started only when the dead time was below 2\,\% minimizing this way the possible error. Additionally, two other methods have been used to check the dead time provided by MAESTRO. In one of the measurements (Zn-3 source) the main amplifier output has been split and the signals have been fed into two completely independent data acquisition systems, the above mentioned ORTEC MAESTRO and a GENIE-2000 system by Canberra \cite{GENIE} which also provides its own dead time values. The half-life analysis has been done independently on the spectra from both systems and the results are included in Table \ref{tab:results} and Fig. \ref{fig:results}. As one can see, the results obtained with the GENIE system are statistically consistent with the MAESTRO result. The difference between the GENIE average and the MAESTRO average of all transitions is 0.02\,h (0.2\,\%).

\begin{figure}
\centering
\resizebox{\textwidth}{!}{{\includegraphics{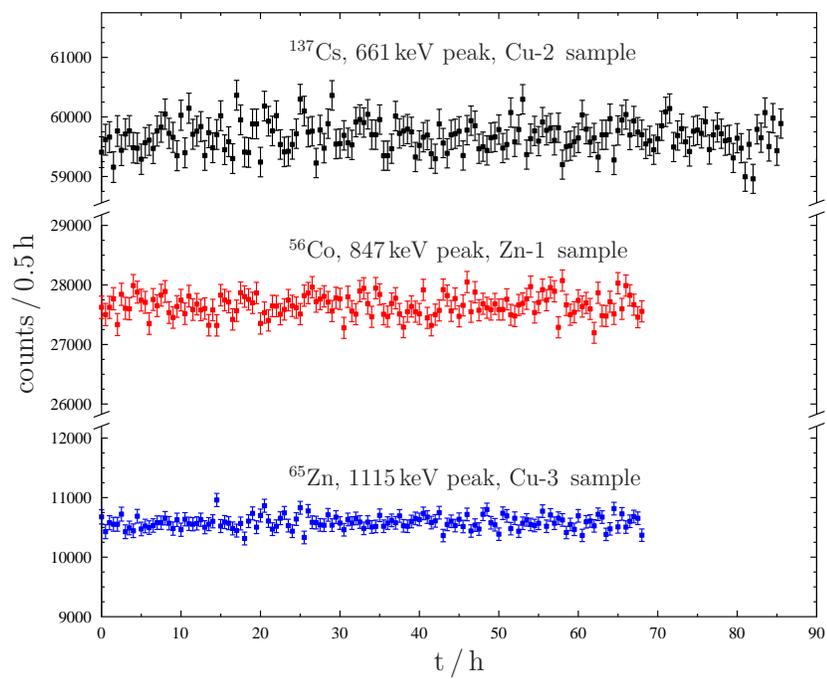}}}
\caption{\label{fig:deadtime} Decay and dead time corrected yield of some of the reference sources used for checking the reliability of the dead time determination.}
\end{figure}

In order to further check the reliability of dead times, long lived reference sources have been added to all counting systems. For the Zn sources $^{56}$Co produced by proton bombardment has been used as the reference source. For the counting of the Cu sources $^{65}$Zn and $^{137}$Cs reference sources have been added. The dead time correction provided by MAESTRO has been applied for the net peak areas of the reference sources and its values as a function of time has been checked. Fig. \ref{fig:deadtime} shows the results in the case of three measurements. The tiny reduction of the reference source activities during the counting period has been corrected for. The figure clearly shows that the decay and dead time corrected yield of the reference sources is stable in time which proves the reliability of the dead time values. In order to quantify this, the yields of the references sources have been fitted with linear functions. The slope of the fitted lines was always consistent with zero. However, regarding the uncertainty of the slopes, a multiplicative correction factor for the dead time values between 0.8 and 1.2 cannot be excluded. (This means that if the dead time provided by MAESTRO is e.g. 2\,\%, then values between 1.6\,\% and 2.4\,\% still lead to a fit of the reference source data consistent with zero slope.) If a multiplicative correction in this range is applied to the $^{66}$Ga half-life analysis, a difference of $\pm$\,0.01\,h is obtained (0.1\,\%). This value is similar to the one obtained from the comparison of the two data acquisition system, therefore the higher of these two values (0.02\,h) is adopted for the systematic uncertainty of the dead time determination.

\subsection{Final half-life result}

Table \ref{tab:uncert} summarizes the uncertainties of the half-life measurement. The statistical uncertainty in the first row was obtained by the average of all individual measurements weighted by their statistical uncertainty given by the exponential fit. For the total uncertainty the quadratic sum of the statistical uncertainty and all the above discussed systematic uncertainties is adopted. The final result of our half-life measurement is therefore t$_{1/2}$\,=\,(9.312\,$\pm$\,0.032)\,h. This value is in very good agreement with the recent high precision experiment (9.304\,h\,$\pm$\,0.008\,h \cite{sev10}) and clearly disagrees with Ref. \cite{abb06} (9.49\,h\,$\pm$\,0.03\,h). Therefore, our new measurement supports the validity of the value of Ref.\,\cite{sev10} from the two recent, strongly contradicting measurements.

\begin{table}
\centering
\caption{\label{tab:uncert} The uncertainty budget of the half-life determination. The total uncertainty is the quadratic sum of all individual uncertainties.}
\begin{tabular}{lcc}
\hline
source & absolute & relative\\
 & uncertainty & uncertainty\\
 \hline
statistics & 0.005\,h & 0.06\,\% \\
source preparation & 0.014\,h & 0.15\,\% \\
detector used & 0.013\,h & 0.16\,\% \\
peak integration & 0.015\,h  & 0.16\,\% \\
dead-time correction & 0.020\,h & 0.21\,\% \\
\hline
total & 0.032\,h & 0.34\,\% \\
\hline
\end{tabular}
\end{table}
 
\section*{Acknowledgements}

This work was supported by the European Research Council StG. 203175 and OTKA grants K68801 and NN83261(EuroGENESIS).

\bibliographystyle{model1a-num-names}
\bibliography{<your-bib-database>}

\end{document}